# Contribution to mid-field Q slope in Niobium SRF cavities from linear RF loss mechanism due to topographic surface structures

C. XU[1,2#], M. J. KELLEY[1,2], C. E. REECE[1]
[1]THOMAS JEFFERSON NATIONAL ACCELERATOR FACILITY, NEWPORT NEWS, VA 23606
[2]COLLEGE OF WILLIAM AND MARY, WILLIAMSBURG, VA 23187

*Abstract*
  Topographic structure on Superconductivity Radio Frequency (SRF) surfaces can contribute additional cavity RF losses describable in terms of surface RF reflectivity and absorption indices of wave scattering theory. At isotropic homogeneous extent, Power Spectrum Density (PSD) of roughness is introduced and quantifies the random surface topographic structure. PSD obtained from different surface treatments of niobium, such Buffered Chemical Polishing (BCP), Electropolishing (EP), Nano-Mechanical Polishing (NMP) and Barrel Centrifugal Polishing (CBP) are compared. A perturbation model is utilized to calculate the additional rough surface RF losses based on PSD statistical analysis. This model will not consider that superconductor becomes normal conducting at fields higher than transition field. One can calculate the RF power dissipation ratio between rough surface and ideal smooth surface within this field range from linear loss mechanisms.

## INTRODUCTION

  RF loss induced by roughness is considered in many RF components, such as micro strip transmission line, wave guide and RF resonator. It can be understood as the RF electromagnetic field penetrates the surface and there the induced current will pass and cause RF loss. [1] However, in a RF wave view, the incident wave is reflected, scattered and absorbed by the rough surface. Inside of a resonator, the reflected, scattered wave contributes to standing wave field, while the absorbed RF wave is attributed to the RF surface loss. These two perspectives may both be used to describe the same RF loss.

  In a resonator, only several specific RF standing wave modes can exist to meet the boundary condition which is the resonator geometry. The electric and magnetic field at one location is combination a of EM components of those plane waves. Within the resonator, E and M are separated in space and interchange their energy over a distance. Thus the peak E and M field are always not the same location. With special EM setup, TE, TM, TEM are used to describe the EM field direction, if presumed direction is beam axis. In some sense, it is very tedious and difficult to expand the field into plane wave expansion. If so, the incident direction should also be from all directions. Therefore, a RF loss calculation method is required and independent of direction. It also covers all frequencies or wavelengths.

## METHODOLOGY

  A rough surface will cost more RF loss. [2] One simple reason is that the surface current have more current path. In another word, the RF wave as more radiation absorption surface. This RF loss will contribute into power consumption and aggravate the quality factor.

  If we consider a 2D random rough surface Z= f(x) in Fig.1. We can expand the magnetic field into Fourier series as in x and z direction. [3]

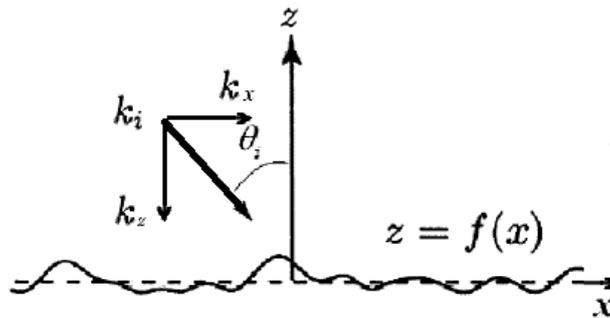

Fig. 1: A plane wave incident impinging on a rough surface with incident angle $\theta_i$.

$$\psi(x,z) = \int_{-\infty}^{\infty} dk_x e^{-jk_x x + jk_{1z} z} \tilde{\psi}(k_x) \qquad (1)$$

Where $k_{1z} = \sqrt{k_1^2 - k_x^2}$ and $k_1 = \dfrac{1-j}{\delta}$. Here δ is the skin depth $\delta = \sqrt{\dfrac{2}{\omega\mu\sigma}}$ and σ is the superconducting conductivity. The physics behind this equation is that the total magnetic field is combination of field component at each spatial wavelength. In another word, the total magnetic field can be expanded into magnetic contribution from each wavelength in spatial frequency.

If we use a second order small perturbation methods, setting

$$\tilde{\psi}(k_x) = \tilde{\psi}^{(0)}(k_x) + \tilde{\psi}^{(1)}(k_x) + \tilde{\psi}^{(2)}(k_x)$$

In first approximation, a fixed constant magnetic field $H_0$ is applied on the surface. Thus, the equation above becomes:

$$H_0 = \int_{-\infty}^{\infty} dk_x \exp(-jk_x x + jk_{1z}xf(x))\tilde{\psi}(k_x)$$

Basically, we have done a Fourier transform to redistribute the magnetic field into each surface spatial wavelength in x direction.

By balancing this equation to second order, we obtain:

$$\tilde{\psi}^{(0)}(k_x) = H_0 \delta(k_x)$$
$$\tilde{\psi}^{(1)}(k_x) = -jk_1 H_0 F(k_x)$$
$$\tilde{\psi}^{(2)}(k_x) = H_0 \int_{-\infty}^{\infty} dk_x' F(k_x - k_x') F(k_x') \left(-k_1 k_{1z}' + \dfrac{k_1^2}{2}\right)$$

For now, we have calculated the RF magnetic field on this given surface. The power absorbed by the conductor, for a given width *w* in *y* direction and length *L* in *x* direction, can be calculated from the Poynting vector.

$$P_a = \dfrac{\omega}{2\sigma} \operatorname{Re} \int ds \dfrac{\partial \psi}{\partial n} \psi^*$$

We simplify the results:

$$<P_a> = \dfrac{\omega L}{2\sigma\delta}\left\{1 + \dfrac{2h^2}{\delta^2}\left[1 - \dfrac{\delta}{h^2}\int_{-\infty}^{\infty} dk_x \left\{k_x W_{1D}(k_x) \operatorname{Re}\sqrt{k_1^2 - k_x^2}\right\}\right]\right\}$$

In order to get rid of the external field, we normalize the power dissipation with that of a smooth surface.

$$\dfrac{<P_a>}{P_{a,smooth}} = 1 + \dfrac{2h^2}{\delta^2} - \dfrac{4\pi}{\delta}\int_0^{\infty} dk_\rho \left\{k_\rho W_{2D}(k_\rho) \operatorname{Re}\sqrt{\dfrac{-2j}{\delta^2} - k_\rho^2}\right\}$$

$W_{2D}(k_\rho)$ is the 2D PSD from an isotropic surface.

Researcher have combined TM and TE wave mode into this equation into a 2D problem. They also avoid the assumption that surface field is a constant.

One should note:

1. The RMS and decay parameters (penetration depth) ratio, $h/\delta$, is critical for characterizing the increased losses.

2. Third term reduces to $h^2/\delta^2$, when the $k_\rho$ is small. The second/third term have net contribution if $1/\delta$ and $k_\rho$ are comparable.

3. Generally, higher $W_{2D}(k_\rho)$ brings additional loss.

## APPLICATION TO SRF SURFACES

To obtain more accurate RF loss ratio, one needs to extend the PSD into as broad a frequency range as possible. Since all characterization method shave cut-off frequencies, one can at most get an extended PSD. The recent extended frequency range is $1/1.2$ cm$^{-1}$–$1/10$ nm$^{-1}$, over 6 decades, with white light interferometry and atomic force microscopy. Because the magnetic field is expanded into horizontal spatial wavelength, the PSD frequency should cover the RF wavelength and beyond. Though an approximation methods is introduced by using Inverse Abel transforms to extend limited range, but obtaining 1D PSD with wider frequency range could improve later calculation accuracy.

---

*Work supported by Jefferson Science Associates. xuchen@jlab.org

$$<W(k_\rho)^{2D}> = -\tfrac{1}{\pi} \times \int_{k_\rho}^{\infty} \frac{dk_\rho}{\sqrt{k_x^2 - k_\rho^2}} \frac{d}{dk_x} <W(k_x)^{1D}>$$

$$<W(k_x)^{1D}> = 2 \times \int_{k_x}^{\infty} \frac{k_\rho dk_\rho}{\sqrt{k_\rho^2 - k_x^2}} <W(k_\rho)^{2D}>$$

This transformation also permits the transformation of the high-frequency behavior of the spectra of one dimensionality to be transformed into the high-frequency behavior of the other without knowledge of their low-frequency behavior.

We investigate SRF surfaces with current polishing methods and materials polished are large/fine/single grain Nb sheets. Buffered chemical polishing, electropolishing and mechanical centrifugal barrel polishing samples are characterized by Atomic Force Microscopy (AFM) and White light interferometry (WLI). Note that these two characterizations have different lateral resolutions and scan scopes which determine the spectral frequency ranges.

1D averaged PSD is calculated by following the routine introduced previously. [4] Such routine includes proper detrending, windowing and averaging.

In this study, the $R_q$ and PSD are used to derive power ratio. Rq value from 4 different locations on each sample is averaged and summarised in Table 1.

Table 1: Table 1: Averaged $R_q$ with AFM and WLI

| Samples | Single crystal | | "Standard" fine-grained | | Centrifugal Barrel Polishing (CBP) | |
|---|---|---|---|---|---|---|
| Treatment | After 30 µm BCP | Nano polished | +100 µm BCP | +50 µm EP | Fine Grain | Large Grain |
| Atomic force microscopy ~5×5µm | | | | | | |
| $R_q$(nm) | 1.99 | 2.84 | 4.89 | 4.40 | 2.76 | 1.20 |
| Atomic force microscopy ~100×100µm | | | | | | |
| $R_q$(nm) | 7.05 | 13.5 | 392.90 | 74.30 | 70.8 | 7.71 |
| White light interferometry ~234×312µm (20x magnification) | | | | | | |
| $R_q$ (nm) | 35.7 | 1.63 | 1500.4 | 291.7 | 151.8 | 13.5 |
| White light interferometry ~930×1244µm (5x magnification) | | | | | | |
| $R_q$ (nm) | 391 | 5.24 | 2008.5 | 642.8 | 335.5 | 49.0 |

One can generalize that Rq value decrease from BCP>EP>CBP Large grain> CBP Fine grain> Single crystal>Nanopolishing samples.

Averaged 1D PSD from AFM/WLI are combined in Fig.2.

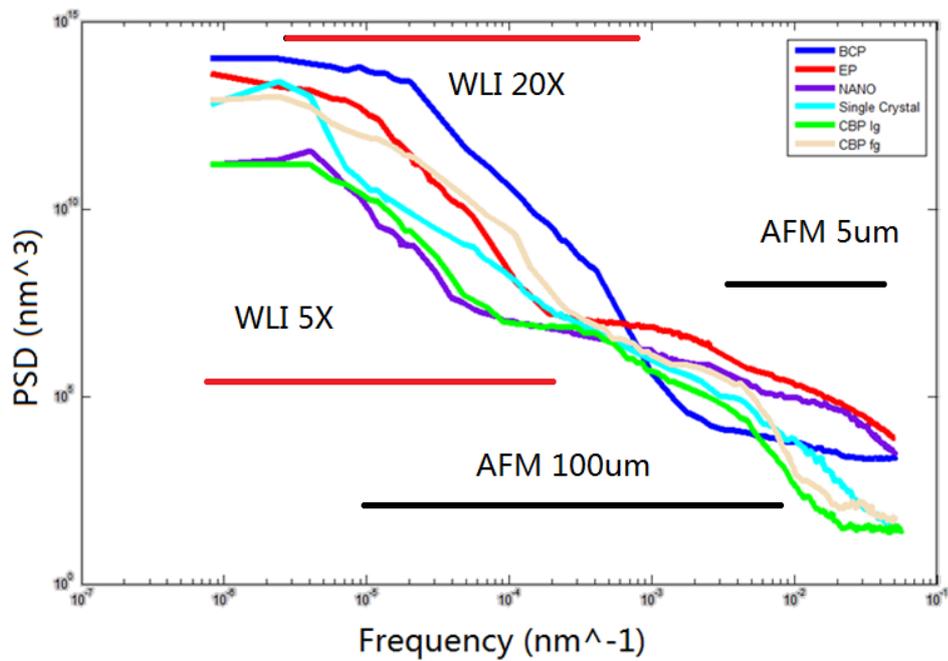

Fig.2: Joint 1D PSD models from AFM/WLI are shown and the characterization frequency domains are indicated.

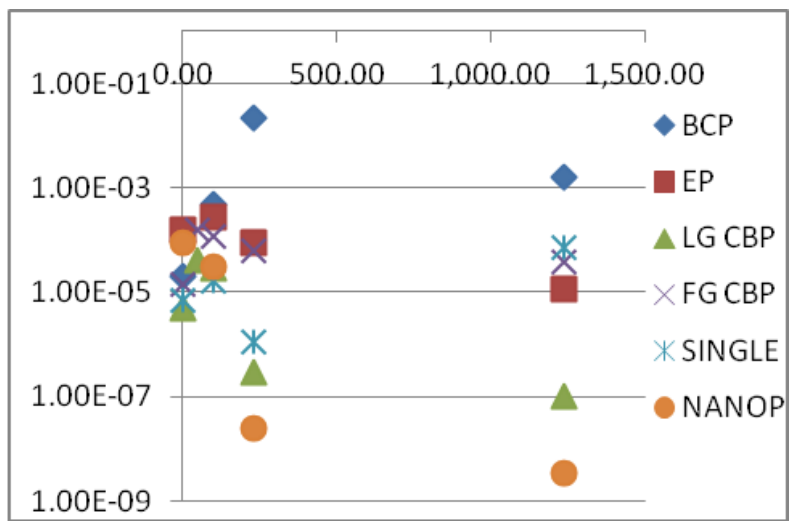

Fig.3 : Power Ratio calculated from 1D PSD. By inverse Abel transform, the 2D PSD are shown in Fig.3.

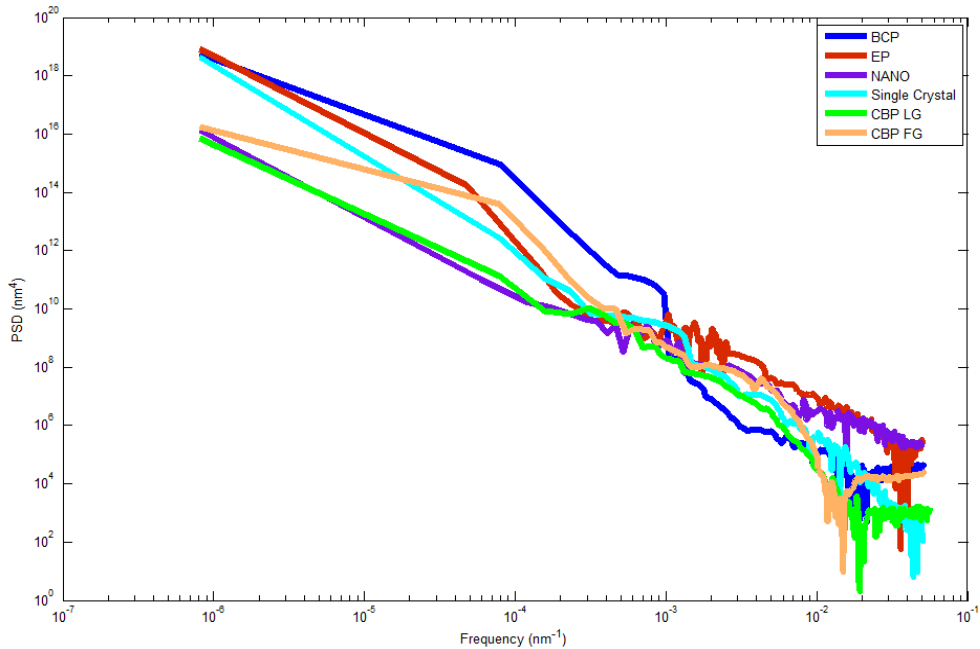

Fig.4: 2D PSD calculated from various surfaces.

Power ratio index are illustrated in Fig.5.

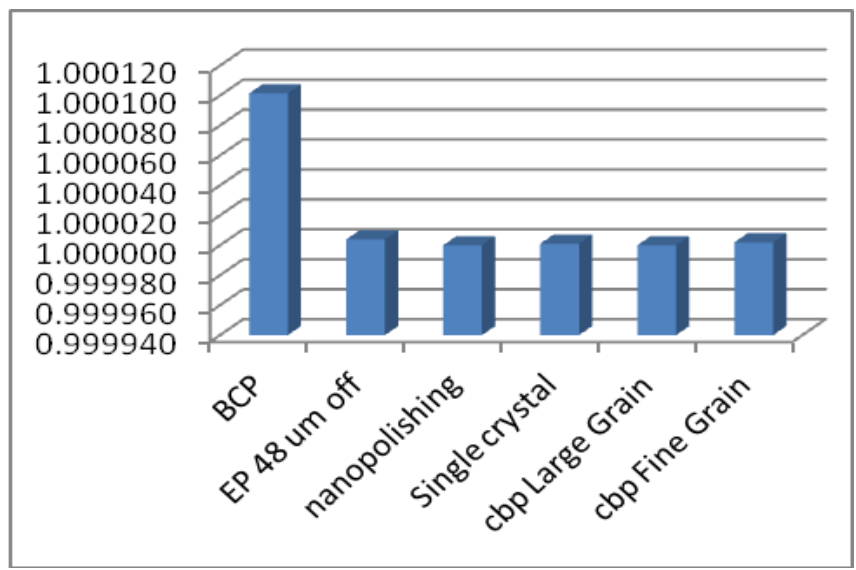

Fig.5 : Power Ratio from 2D isotropic PSD.

BCP surface show higher ratio, but other surfaces suggest almost the same closed to 1.

## DISCUSSION

One can see from power ratio equation that the last term is an integration of 2d PSD.

In equation of power ratio, without the Re(…), the integrand is simply the square of RMS height. One can infer that if the $k_\rho$ is small enough compared with $2/\delta$, then the Re(…) term reduced into $1/\delta$. In that limit the second and third terms cancel each other. The total power ratio becomes one. This gives substantiates the interpretation that features at larger wavelength have less RF power loss than the small high frequency features. Another understanding is that only features with lateral extent comparable to the penetration length give a significant effect on **the additional power loss ratio.** Applying this analysis to variously prepared niobium surfaces typical of those in SRF cavities, we find that linear RF losses depend negligibly on roughness for any of the characteristic surfaces considered. On the other hand, SRF materials are particularly susceptible to non-linear and temperature-dependent losses. The non-linear losses are reflected in the observed *Q* drop with increasing surface magnetic field. We are examining the influence of topography on such losses separately [5].

One of advantages of SRF Cavity technology is that the quality factor is much higher than the copper cavities. This mainly is due to the fact that the surface resistance is much smaller than the conventional normal conducting cavities. This surface resistance reflects the RF power absorbed and consumed on the surface. Due to the superconductivity, magnetic field only penetrates into surface within narrow depth. Thus very outer surface contributes the most RF consumption. Surface topography is the most direct factor that intuitively affects surface resistance. People use several polishing methods to achieve smooth surface at reasonable cost. However, there are no definitive trends indicating that the different polishing methods have systematic surface resistance priority than the others in term of cavity Quality factor. This study aims to answer the puzzle how additional power loss attributed from surface roughness by adapting the electromagnetic wave scattering theory. This Hyungen scattering method is equivalent as solving Maxwell equation for EM fields. A statistical analysis is given in form of spectrum to describe surface roughness in the frequency domain. With assumption that surface is isotropic, this 1D spectrum can be expanded into 2D spectrum in order to study the state of art surfaces polishing technologies. A power ratio between rough surfaces over the flat surface is given, and this index is only surface geometry related. Index from surfaces with different treatments are compared and discussed in details. There is nonlinearity relation RF loss and harmonic lateral scale and one can calculate RF loss within a period of frequency. With characteristic spectrum of each surface treatment, power loss ratio is given at different frequency domains for each treatment.

One can see the 1D PSD from different surface treatments have a crossover at 1um-1. This is mainly because the grain size. As we claimed before, BCP and EP surface modify surface differently at intra grain and inter grains scales. BCP polishing intra grain surface but differentiate inter grain, vice versa, EP smoothen inter grains but roughen surface inside of grain. Accordingly, 2D PSD show the similar trends that the crossover is also at 1um.

Carefully compare the PSD from single crystal and CBP samples; one can see that three types surfaces have almost the same PSD amplitude beyond 1um. This means the CBP technology basically didn't change surface roughness within this range. However, under 1um frequency, CBPed fine grain has the highest value, and single crystal and CBPed large grain surface are almost the same. This suggests that inter grain roughness contributes in this range and CBP could not overcome this inter grain roughness harmonica.

Corresponding this with the power ratio, even though the ration absolute number is very small, we find ratio values follow the same PSD amplitude trend. The ratio indexes have a crossover at frequency 25um-1. The indexes for EP remains the same while BCP increases and reaches peak at 200um and steady reduces. Index from nanopolishing surface basically decreases as the flat surface, because the harmonica with lateral length comparable to penetration depth has contribution. Other surfaces all show a little peak at frequency 30um-1. These indexes are calculated from the characterization where frequency range may overlap.

~~Another study is promoted and power ratio is calculated from different frequency regions from 2D PSD. Simply take 1um-1 as a boundary,~~

## 5. CONCLUSION

Spectral study suggested that surface roughness doesn't contribute much in the linear RF loss, because current practical surface a smooth enough at a wavelength closed to penetration depth.

## ACKNOWLEDGMENT

Authored by Jefferson Science Associates, LLC under U.S. DOE Contract No. DE-AC05-06OR23177. The U.S. Government retains a non-exclusive, paid-up, irrevocable, world-wide license to publish or reproduce this manuscript for U.S. Government purposes.